\documentclass[conference,final]{IEEEtran}
\IEEEoverridecommandlockouts
\usepackage{cite}
\usepackage{amsmath,amssymb,amsfonts}
\usepackage{algorithmic}
\usepackage{graphicx}
\usepackage{textcomp}
\usepackage{algorithm}
\usepackage{CJKutf8}
\usepackage[encapsulated]{CJK}
\usepackage{graphicx}
\usepackage{float}

\newtheorem{theorem}{Problem}

\newcommand{\eg}{\emph{e.g.}}

\def\BibTeX{{\rm B\kern-.05em{\sc i\kern-.025em b}\kern-.08em
    T\kern-.1667em\lower.7ex\hbox{E}\kern-.125emX}}
\begin{document}

\title{Computing Lens for Exploring the Historical People's Social Network
}

\author{\IEEEauthorblockN{Junjie Huang}
\IEEEauthorblockA{\textit{University of Chinese Academy of Sciences}\\
Bejing, China \\
huangjunjie17@mails.ucas.ac.cn}
\and
\IEEEauthorblockN{Tiejian Luo}
\IEEEauthorblockA{\textit{University of Chinese Academy of Sciences}\\
Beijing, China \\
tjluo@ucas.ac.cn}
}

\maketitle

\begin{abstract}
A typical social research topic is to figure out the influential people's relationship and its weights. 
It is very tedious for social scientists to solve those problems by studying massive literature. Digital humanities bring a new way to a social subject. 
In this paper, we propose a framework for social scientists to find out ancient figures' power and their camp. 
The core of our framework consists of signed graph model and novel group partition algorithm. 
We validate and verify our solution by China Biographical Database Project (CBDB) dataset.
The analytic results on a case study demonstrate the effectiveness of our framework, which gets information that consists with the literature's facts and social scientists' viewpoints.
\end{abstract}


\section{Introduction}
As we all know, microscopes can help bioscientists better observe the internal structure of cells.
Our framework hopes to help researchers not only focus on the relationships of specific research characters but recognize the role of these characters in the entire large social network, which we hope it works like computing lens for some social problems.

The traditional method to study historical persons' relationship is mainly through reading literature, accessing a large number of historical documents and demonstrating the relationships.
And it is mostly qualitative analysis.
However, this method requires too much professional knowledge. 
For example, for someone who does not understand the ancient literature, it is impossible to dig out information in historical documents.
In addition, the traditional method used to discuss the relationship between two people was limited to the direct relationship between historical people but did not consider the influence of these historical people's friends and even the entire social network on him.
Therefore, this paper proposes a new social network research framework(CLHPSoNet) to help people understand the influence of social networks of historical people on the social relationships of historical people.
This method can be studied quantitatively without the need for rich background knowledge.
The comparison between our framework and traditional methods are shown in Table \ref{tb:comparison}.
Our method can get some answers about the problems like who is the central figures? or what's the relationship between some people?\footnote{https://www.zhihu.com/question/20589740}.
These questions are not only concerned by researchers, but history enthusiasts also have great interests.

 \begin{table}[h]
\centering
\caption{Comparison of our framework and traditional methods}
\label{tb:comparison}

\begin{tabular}{lcc} 
\hline
    & Traditional methods   & Our framework\\ 
\hline
Data source     & Historical Documents & Database     \\
Research methods& Qualitative   & Quantitative \\
\begin{tabular}[c]{@{}l@{}}Relationship between~\\research objects\end{tabular} & Direct Relationship & Network  \\
\begin{tabular}[c]{@{}l@{}}Expert knowledge\\requirements\end{tabular} & Professional  & Normal\\
Results display & Papers, Tables& Web UI\\
\hline
\end{tabular}
\end{table}

The major contributions of this paper are as follows:
\begin{itemize}
    \item A novel research framework for exploring the social relationship of historical people is presented. 
    \item We propose graph partitioning algorithms to solve relevant domain problems.
    \item Based on the proposed model and algorithm, we have built an application to help people analyze and understand the social relationships of the ancients.
\end{itemize}

The rest of the paper is organized as follows.
We introduce related work in Section 2.
We introduce our modeling in Section 3 and system designing in Section 4.
An Application is built in Section 5 to show our framework work well in some domain problems.
We conclude the paper and point out future directions in Section 6.


\section{Related Work}
\subsection{Research on Historical Figures' Relationship}
The study of social relationships between historical people plays an important role in the study of history.
In addition to a richer account of the life experiences of the characters, it also reflects the historical background of the historical people to some extent. 
It attracted a lot of research attention.

For example, Liu \cite{liunaichang1981} lists historical document records to discuss the Su Shi’s\footnote{https://en.wikipedia.org/wiki/Su\_Shi}(a song dynasty poet) association with Wang Anshi\footnote{https://en.wikipedia.org/wiki/Wang\_Anshi} (a song dynasty poet), from life experience, political, literature, and other aspects.
Guo\cite{guyongxin2001} use a lot of historical documents to discuss the friendship between Ouyang Xiu and Wang Anshi. 
With the development of digital humanities, some researchers input the historical document records to computers to count, compute and visualize the relationship of historical people.
Yu et al\cite{yushihua2013} have some statistics on the friendship between Su Shi and Wang Gong(a song dynasty poet).
Zhu \cite{zhu2016chro} proposed a `Chro-Ring' approach to visualize the history of poets.
However, most of the research in this area is qualitative, straightforward and historical-based.

\subsection{Research on China Biographical Database Project (CBDB)}
The China Biographical Database(CBDB)\footnote{https://projects.iq.harvard.edu/cbdb}\cite{fuller2015china} is a freely accessible relational database with biographical information about approximately 417,000 individuals, primarily from the 7th through 19th centuries. 
It is developed by Harvard University, Institute of History and Philology of Academia Sinica and Peking University. 
Recent CBDB's version released in April 2017, it has been continuously updated.
The data is meant to be useful for statistical, social network, and spatial analysis as well as serving as a kind of biographical reference.

Based on CBDB dataset, a lot of new applications and system are proposed. 
Liu et al\cite{liu2016tracking, liu2017flexible} use the database to compare the poetry of different dynasties, it shows the accessibility of increasingly larger datasets strengthens researchers' research potential.
Guo\cite{guoxiuping2016} proposed a powerful tool for studying genealogical records. 
Sie et al\cite{sie2017development}  developed a text retrieval and mining system for Taiwanese historical people.
There exist lots of valuable information waiting for researchers to mine.
Our paper uses the dataset to implement our framework. 
Of course, other historical files or database can be added to make it more complete.

\subsection{Signed Graphs}

Signed social networks are such social networks in which social ties can have two signs: positive and negative\cite{leskovec2010signed}.
It was first addressed by social balance theory, which originated in cognitive theory from the 1950s\cite{heider1946attitudes}.
For all signed triad, triads with an odd number of positive edges are defined as balanced.
It was proven that if a connected network is balanced, it can be split into two opposing groups and vice versa\cite{easley2010networks}.
In reality, however, social networks are rarely, structural balance theory has been developed extensively, e.g. weak structural balance, k-balance, and other different criteria to quantify and evaluate balance in a signed social network\cite{levorato2015ils, leskovec2010signed, srinivasan2011local}.
Researchers defined clustering problems on signed graphs can be used as a criteria to measure the degree of balance in social networks\cite{doreian1996partitioning}.
Then it becomes a cluster problem and it's NP-hard, a lot of heuristic methods are proposed to solve the clustering problems\cite{doreian2009partitioning,traag2009community ,doreian1996partitioning, levorato2015ils}.
Recently, with the development of network representation learning, researchers begin to use machine learning methods to learn low-dimensional vector representations for nodes of a given network.
And it has been proven to be useful in many tasks of network analysis such as link prediction\cite{wang2017signed}, node classification\cite{grover2016node2vec}, and visualization\cite{maaten2008visualizing}.
However, network embedding has various methods (e.g. DeepWalk\cite{perozzi2014deepwalk}, node2vec\cite{grover2016node2vec}, t-SNE\cite{maaten2008visualizing})been proposed for unsigned network embedding. How to do network embedding for signed graph attracted the interest of researchers. 
A framework like SiNE\cite{wang2017signed} is the latest attempt, it inspires the use of embedding methods for sigend graph.

\section{Modeling}

\subsection{Data Preparation}
As we said in related work, we use the CBDB's API\footnote{https://projects.iq.harvard.edu/cbdb/cbdb-api} to collect data. We cleaned the collected data. 
According to the following rules, we divided the dynasty where people belong to:
\begin{enumerate}
    \item The dynasty is marked
    \item The year of birth is within the year of the dynasty
    \item The year of death is within the year of the dynasty
\end{enumerate}

\begin{figure*}[!h]
    \centering
    \includegraphics[width=\textwidth]{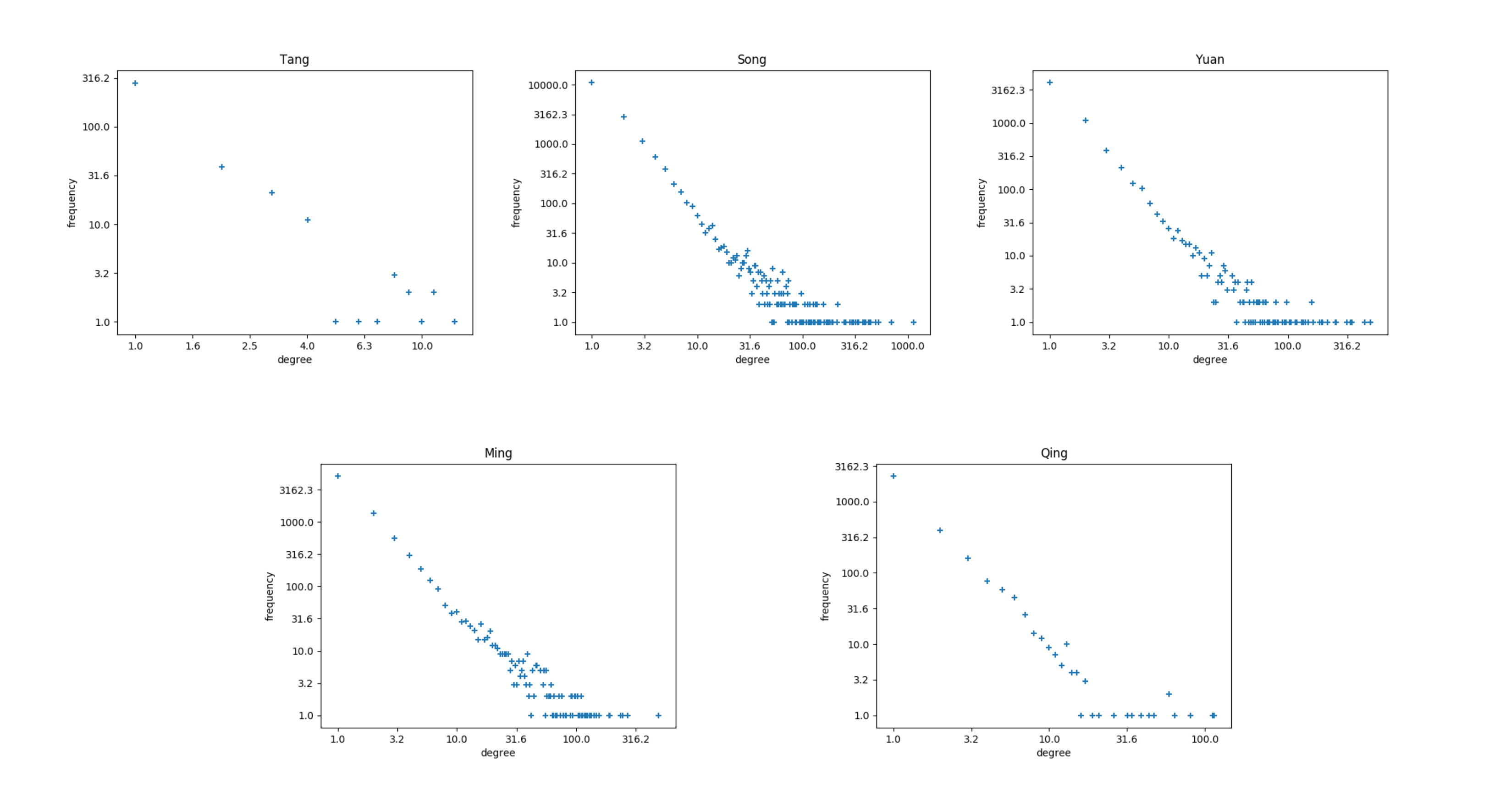}
    \caption{Logarithmic plot of the degree distribution showing that the degree distribution in the networks follows a power law.}
    \label{fig:networks-all}
    
\end{figure*}

We use NetworkX\footnote{https://networkx.github.io/} to do some statistics on the networks, which the number of nodes $|V|$, the number of edges $|E|$, average clustering coefficient\cite{saramaki2007generalizations} and average path length. 
The results show in the Table \ref{tb:networks}. 
Besides, Figure \ref{fig:networks-all} shows the degree distributions in networks. 
It shows that the social network in Tang dynasty is not complete. 
Other networks show a power law phenomenon and small world phenomenon. 
For our following works, we use Song dynasty social networks.

\begin{figure}[!htb]
    \centering
    \includegraphics[height=0.35\textwidth]{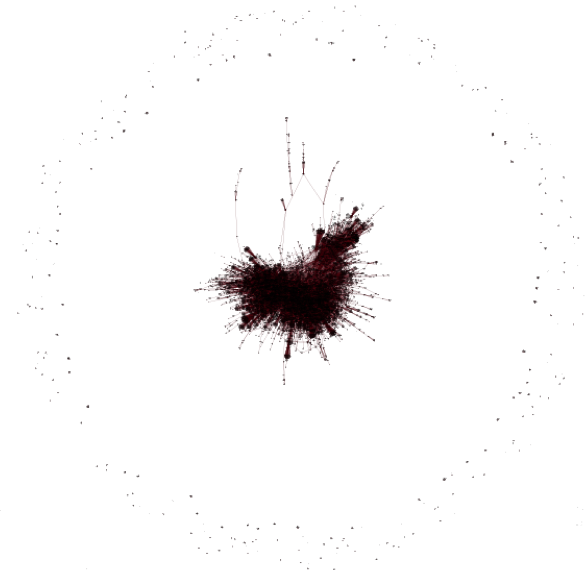}
    \caption{The social network in Song dynasty is visualized, it shows that the network consists of a large core component and other marginal components. }
    \label{fig:song-all}
\end{figure}

\begin{figure}[!htb]
    \centering
    \includegraphics[height=0.35\textwidth]{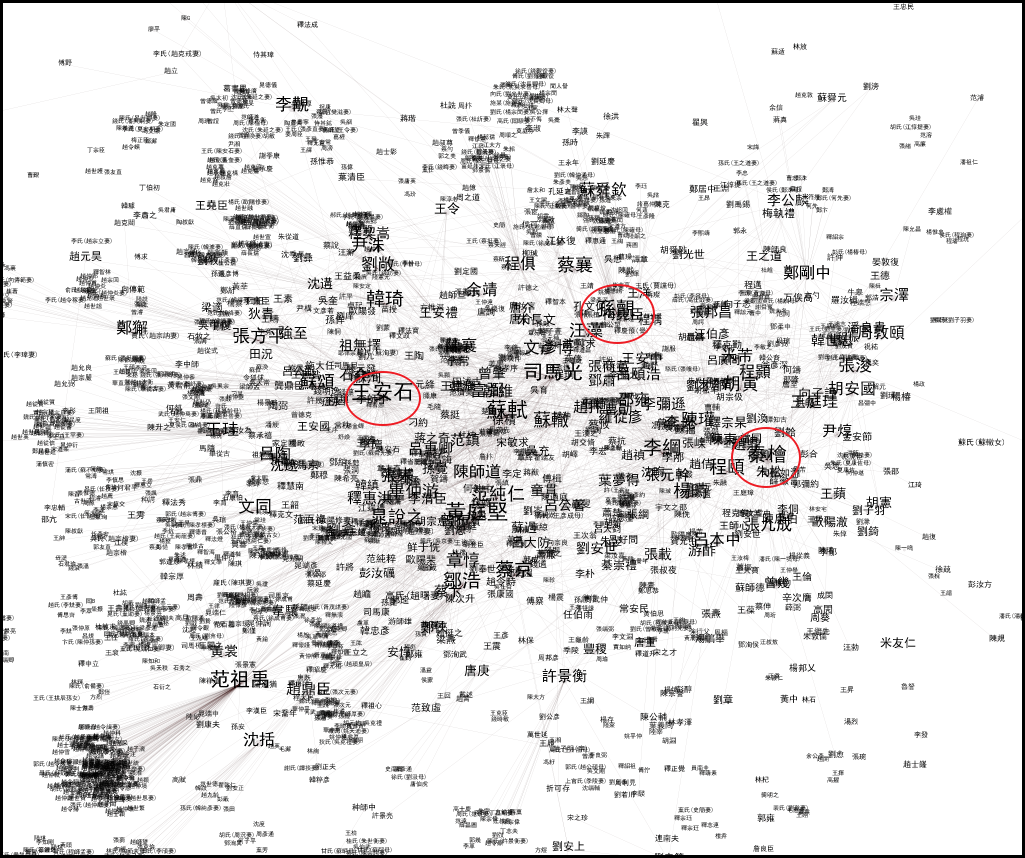}
    \caption{Wang Anshi (1021-1086) as the focus, the relationship between people born in 1000-1100. It shows that Wan Anshi can get a relationship with somebody like Qing Hui (1091-1155)}
    \label{fig:wanganshi}
\end{figure}

\begin{table}[h]
\centering
\caption{Network Information of Different Dynasty}
\label{tb:networks}
\begin{tabular}{lccccc}
\hline
     & $|\mathcal{V}|$  & $|\mathcal{E}|$   &  \begin{tabular}[c]{@{}l@{}}Average \\Clustering \\ Coefficient\end{tabular} & \begin{tabular}[c]{@{}l@{}}Average \\Path \\ Length\end{tabular} \\ \hline
Tang(618, 907) & 365   & 286   & 0.016                  & 1.60                \\
Song(960, 1279) & 17,114 & 30,330 & 0.121                  & 4.08                \\
Yuan(1271, 1368) & 6,424  & 11,864 & 0.150                  & 4.00               \\
Ming(1368, 1644) & 8,350  & 14,609 & 0.070                  & 4.65                \\
Qing(1636, 1912) & 3,128  & 3,059  & 0.021                  & 7.71                \\ \hline
\end{tabular}
\end{table}
In order to get an intuitive impression, we use Gephi\footnote{https://gephi.org/} to visualize the network to get Figure \ref{fig:song-all} and Figure \ref{fig:wanganshi}. 
It shows that most of the important people are contained in a big connected component, which shows a small world phenomenon. 
And if you focus on someone \eg Wan Anshi, you will find that even someone \eg Hui Qin\footnote{https://en.wikipedia.org/wiki/Qin\_Hui} whom Wan Anshi don't know can be reached by little steps, such as Sun Di.

\subsection{Domain Problem}

After our data preparation, we have a whole undirected weighted graph $G = (V, E, w)$. $w_{ij}$ is the number of records between $v_i$ and $v_j$.
And we want to solve some problems people concerned. 
For example, how is the relationship between Eight Great Prose Masters of the Tang and Song\footnote{https://en.wikipedia.org/wiki/Eight\_Masters\_of\_the\_Tang\_and\_Song}.
Or who is the central figures in the New Policies in Song dynasty\footnote{https://en.wikipedia.org/wiki/New\_Policies\_(Song\_dynasty)}?

To answer the questions about the relationship for some people, we need do some data modelings.

\subsection{Data Model}

\subsubsection{signed networks}
As we mentioned in the related work about the signed graph. 
In order to answer some domain problems, the network needs to be a signed graph to make the relationship more semantics, which means to make the relationship friendly or unfriendly. 
We assigns a sign to each edge in $E$, then $G = (V,E,w,s)$ is achieved. 
A total 445 kinds of relationship is manually annotated. 
Some signed rules(Top 10) showed in the Table \ref{tb:signed-rule}. 
It shows that the number of positive relationships is much more than the number of negative relationships.
\begin{table*}[!htb]
    \centering
    \caption{the statistics of top10 signed rules }
    \label{tb:signed-rule}
\scalebox{0.8}{
    \begin{tabular}{p{2.5cm}p{2.5cm}p{1cm}p{2.5cm}p{2.5cm}p{1cm}p{2.5cm}p{2.5cm}p{1cm}} 
    \hline
    \multicolumn{3}{c}{Positive}& \multicolumn{3}{c}{Negative}       & \multicolumn{3}{c}{Neutral}                    \\ 
    \hline
    Relationship(Chinese)               & Relationship& \multicolumn{1}{l}{Counts} & Relationship(Chinese)                  & Relationship       & \multicolumn{1}{l}{Counts} & Relationship(Chinese)                & Relationship                   & \multicolumn{1}{l}{Counts}  \\ 
    \hline
    \begin{CJK*}{UTF8}{gbsn}为\end{CJK*}Y\begin{CJK*}{UTF8}{gbsn}作墓志铭\end{CJK*} & Make epitaph for Y                & 10189& \begin{CJK*}{UTF8}{gbsn}弹劾\end{CJK*}     & impeach            & 400  & \begin{CJK*}{UTF8}{gbsn}被致书由\end{CJK*}Y & The book was written by Y      & 3624  \\
    \begin{CJK*}{UTF8}{gbsn}墓志铭由\end{CJK*}Y\begin{CJK*}{UTF8}{gbsn}所作\end{CJK*}& Epitaph made by Y                 & 8502 & \begin{CJK*}{UTF8}{gbsn}被\end{CJK*}Y\begin{CJK*}{UTF8}{gbsn}弹劾\end{CJK*}   & Invoked by Y       & 385  & \begin{CJK*}{UTF8}{gbsn}致书\end{CJK*}Y   & mail to Y& 3302  \\
    \begin{CJK*}{UTF8}{gbsn}书序由\end{CJK*}Y\begin{CJK*}{UTF8}{gbsn}所作\end{CJK*} & Book sequence made by Y           & 4774 & \begin{CJK*}{UTF8}{gbsn}遭到\end{CJK*}Y\begin{CJK*}{UTF8}{gbsn}的反对\end{CJK*}/\begin{CJK*}{UTF8}{gbsn}攻讦\end{CJK*}                    & Y's opposition/attack                    & 348  & \begin{CJK*}{UTF8}{gbsn}同僚\end{CJK*}    & A colleag& 538   \\
    \begin{CJK*}{UTF8}{gbsn}为\end{CJK*}Y\begin{CJK*}{UTF8}{gbsn}所著书作序\end{CJK*}& Preface to Y's book               & 4323 & \begin{CJK*}{UTF8}{gbsn}反对\end{CJK*}/\begin{CJK*}{UTF8}{gbsn}攻讦\end{CJK*}  & Opposition/attack  & 334  & \begin{CJK*}{UTF8}{gbsn}未详\end{CJK*}    & Unknown  & 353   \\
    \begin{CJK*}{UTF8}{gbsn}书跋由\end{CJK*}Y\begin{CJK*}{UTF8}{gbsn}所作\end{CJK*} & Book made by Y                    & 4111 & \begin{CJK*}{UTF8}{gbsn}反对\end{CJK*}/\begin{CJK*}{UTF8}{gbsn}不支持\end{CJK*}Y\begin{CJK*}{UTF8}{gbsn}的政策\end{CJK*}                   & Oppose/Do not support Y's policy         & 297  & (\begin{CJK*}{UTF8}{gbsn}暂时保留，待删除：吏部供职\end{CJK*})             & (Temporarily reserved pending deletion: Appointment) & 331   \\
    \begin{CJK*}{UTF8}{gbsn}祭文由\end{CJK*}Y\begin{CJK*}{UTF8}{gbsn}所作\end{CJK*} & The memorial was made by Y        & 3975 & \begin{CJK*}{UTF8}{gbsn}其政策被\end{CJK*}Y\begin{CJK*}{UTF8}{gbsn}反对\end{CJK*}/\begin{CJK*}{UTF8}{gbsn}不支持\end{CJK*}                  & Its policy is opposed by Y/not supported & 281  & \begin{CJK*}{UTF8}{gbsn}上司为\end{CJK*}Y  & Boss is Y& 213   \\
    \begin{CJK*}{UTF8}{gbsn}为\end{CJK*}Y\begin{CJK*}{UTF8}{gbsn}所著书作跋\end{CJK*}& Writing for Y's book              & 3868 & \begin{CJK*}{UTF8}{gbsn}遭\end{CJK*}Y\begin{CJK*}{UTF8}{gbsn}排挤\end{CJK*}   & Excluded by Y      & 199  & \begin{CJK*}{UTF8}{gbsn}为\end{CJK*}Y\begin{CJK*}{UTF8}{gbsn}之考官\end{CJK*} & The examiner for Y             & 167   \\
    \begin{CJK*}{UTF8}{gbsn}为\end{CJK*}Y\begin{CJK*}{UTF8}{gbsn}作祭文\end{CJK*}  & Make a memorial for Y             & 3667 & \begin{CJK*}{UTF8}{gbsn}排挤\end{CJK*}     & Exclude            & 196  & \begin{CJK*}{UTF8}{gbsn}奏录\end{CJK*}Y\begin{CJK*}{UTF8}{gbsn}之文\end{CJK*} & Record the story of Y          & 20    \\
    \begin{CJK*}{UTF8}{gbsn}为\end{CJK*}Y\begin{CJK*}{UTF8}{gbsn}作临別赠言\end{CJK*}(\begin{CJK*}{UTF8}{gbsn}送別诗、序\end{CJK*})               & Make a parting speech for Y (Farewell poetry            & 3623 & \begin{CJK*}{UTF8}{gbsn}得罪\end{CJK*}Y    & Offend Y           & 177  & \begin{CJK*}{UTF8}{gbsn}同场屋\end{CJK*}/\begin{CJK*}{UTF8}{gbsn}同应举\end{CJK*}                     & Same-room housing              & 20    \\
    \begin{CJK*}{UTF8}{gbsn}临別得到\end{CJK*}Y\begin{CJK*}{UTF8}{gbsn}所作赠言\end{CJK*}(\begin{CJK*}{UTF8}{gbsn}送別诗、序\end{CJK*})             & Make a goodbye to Y's speech (Farewell poems, prefaces) & 3520 & \begin{CJK*}{UTF8}{gbsn}不合\end{CJK*}     & Not fit            & 152  & \begin{CJK*}{UTF8}{gbsn}以宦官事\end{CJK*}Y & With official business Y       & 18    \\
    \hline
    \end{tabular}
}
    \end{table*}

\subsubsection{Top and Central People}
Centrality is a common way to quantify the importance of the vertices in a network. 
It has a lot of different declinations are used in social network analysis to identify and highlight specific nodes \cite{grandjean2016social}. 
In order to find the important people in our network, we use the \textbf{Degree Centrality, Betweenness Centrality, Closeness Centrality and Eigenvector Centrality}\cite{wasserman1994social}. 
In Table \ref{tb:central-people}, Top 15 degree centrality figures are listed, which is order by \textbf{Degree Centrality}. 
It's easy to find that most high degree centrality people are famous in the history. 

\begin{table}[H]
    \centering
    \caption{The centrality of Top 15 People order by Degree Centrality}
    \label{tb:central-people}
    \scalebox{0.8}{
        \begin{tabular}{lccccc}
        \hline
        EngName        & PersonId & \begin{tabular}[c]{@{}l@{}}Degree\\ Centrality\end{tabular} & \begin{tabular}[c]{@{}l@{}}Betweenness\\  Centrality\end{tabular} & \begin{tabular}[c]{@{}l@{}}Closeness \\ Centrality\end{tabular} & \begin{tabular}[c]{@{}l@{}}Eigenvector \\ Centrality\end{tabular} \\ \hline
        Zhu Xi         & 3257     & 0.066           & 0.170                  & 0.399               & 0.400                   \\
        Unknown        & 9999     & 0.041           & 0.065                 & 0.341               & 0.191                 \\
        Zhou Bida      & 7197     & 0.031           & 0.061                 & 0.359               & 0.203                 \\
        Su Shi         & 3767     & 0.029           & 0.083                 & 0.380                & 0.179                 \\
        Wu Cheng       & 10084    & 0.026           & 0.055                 & 0.345               & 0.072                 \\
        Liu Kezhuang   & 3595     & 0.026           & 0.055                 & 0.342               & 0.105                 \\
        Wang Anshi     & 1762     & 0.025           & 0.054                 & 0.361               & 0.138                 \\
        Wei Liaoweng   & 4001     & 0.025           & 0.054                 & 0.356               & 0.131                 \\
        Lou Yue        & 3624     & 0.023           & 0.042                 & 0.348               & 0.139                 \\
        Huang Tingjian & 7111     & 0.023           & 0.049                 & 0.363               & 0.132                 \\
        Ouyang Xiu     & 1384     & 0.022           & 0.046                 & 0.362               & 0.135                 \\
        Lv Zuqian      & 7055     & 0.020            & 0.027                 & 0.334               & 0.111                 \\
        Yu Ji(2)       & 10801    & 0.020           & 0.047                 & 0.346               & 0.073                 \\
        Wang Yun(5)    & 28617    & 0.020            & 0.022                 & 0.321               & 0.048                 \\ \hline
        \end{tabular}
    }
\end{table}

\textbf{Degree Centrality} defines the number of links that have a node. For a graph, $G = (V, E)$, the degree centrality for vertex $v$ is:
$$C_d(v) = \frac{deg(v)}{|V|-1}$$
$deg(v)$ is the degree of $v$. The degree centrality values are normalized by dividing by the maximum possible degree $|V|-1$ in a simple graph. 
It shows that people who know many other people are actually important and powerful. \eg Anshi Wang who once served as prime minister and implemented reforms New Policies\footnote{https://en.wikipedia.org/wiki/New\_Policies\_(Song\_dynasty)}.
This reform involves many aspects, such as finance, military, education an so on. 
As the reformer, Wang Anshi were in the political center and contacted with many people.
What's more, in the network, the distribution of `long tail' of this measure exists like many other real-world network\cite{grandjean2016social}, which means a lot of people have a low degree centrality. 

\textbf{Betweenness Centrality} is a centrality measure of a vertex within a graph. 
Vertices that occur on many shortest paths between other vertices have higher betweenness than those that do not. 
The betweenness computed as follows:
$$C_b(v) = \sum_{s,t \in V} \frac{\sigma(s,t|v)}{\sigma(s,t)}$$
where $\sigma(s,t)$ the number of shortest $(s,t)$-paths, $\sigma(s,t|v)$ is the number of those paths passing through some node $v$ other than $s,t$. 
It measures the person's ability to act as a bridge between other people.
In our networks, some people with high degree centrality may not have high betweenness centrality like Lv Zuqian.

\textbf{Closeness Centrality} is the reciprocal of the average shortest path distance to  over all other reachable nodes. 
The result is “a ratio of the fraction of actors in the group who are reachable, to the average distance” from the reachable actors. 
It defines as follows:
$$C_c{v} = \frac{n-1}{|V|-1}\frac{n-1}{\sum_{v=1}^{n-1} d(v,u)}$$
where $d(v, u)$ is the shortest-path distance between $v$ and $u$, and $n$ is the number of nodes that can reach $u$.
It can measure someone's independence, influence and the ability of information transmission.

\textbf{Eigenvector Centrality} computes the centrality for a node based on the centrality of its neighbors.
$$A x = \lambda x$$
where $A$ is the adjacency matrix of the graph $G$ with eigenvalue $\lambda$. The power iteration method is used to compute the eigenvector, the default is 100 iterations. 
Eigenvector centrality assigned to the vertices according to the score their neighbors received.
It takes the number of people and the power of the people someone knows into considerations.

\subsubsection{Subgraph Extraction}
In order to measure the social balance, we proposed an Algorithm \ref{alg:algorithm1}. If $d = 0$, the graph will be the direct relationship between the people we care about.  
Algorithm \ref{alg:algorithm1} can extract the subgraph centered on the seed node, which can better reflect the relationship between seed nodes.
Because of small world phenomenon, $d$ should less than 4 most cases and the number of nodes $|V_{sub}|$ in subgraph grow very fast.

\begin{algorithm}[t]
    \caption{Extract Subgraph}
    \label{alg:algorithm1}
    \begin{algorithmic}[1]
    \renewcommand{\algorithmicrequire}{\textbf{Input:}}
    \renewcommand{\algorithmicensure}{\textbf{Output:}}
    \REQUIRE  seed people $V_{seed} = \{v_1, v_2, ..., v_n\}$, depth $d$, graph $G = \{V , E, w, s\}$
    \ENSURE  subgraph $G_{sub}$
    \\ \textit{Initialisation} : $V_{sub}$ = $V_{seed}$,$k = d, currentSet = V_{seed}$
    \WHILE{ $k > 0$}
    \STATE{$nextNodeSet = \emptyset, k = k - 1$}
        \FOR{$n$ in $currentSet$}
            \STATE{neighborSet = getNeighbors($n$)}
            \FOR{$n_2$ in neighborSet}
                \IF {$n_2 \notin V_{sub}$}
                \STATE {$nextNodeSet$ = $nextNodeSet \cup \{n_2\}$}
                \ENDIF
            \ENDFOR
        \ENDFOR
    \STATE{$V_{sub} = V_{sub}\cup nextNodeSet$}
    \STATE{$currentSet=nextNodeSet$}
    \ENDWHILE
    \STATE $G_{sub} = getSubGraph(V_{sub})$
    \RETURN $G_{sub}$
\end{algorithmic} 
\end{algorithm}

\section{System Design}

\subsection{Problem Definition}

\begin{figure*}[!ht]
    \centering
    \includegraphics[width=0.8\textwidth]{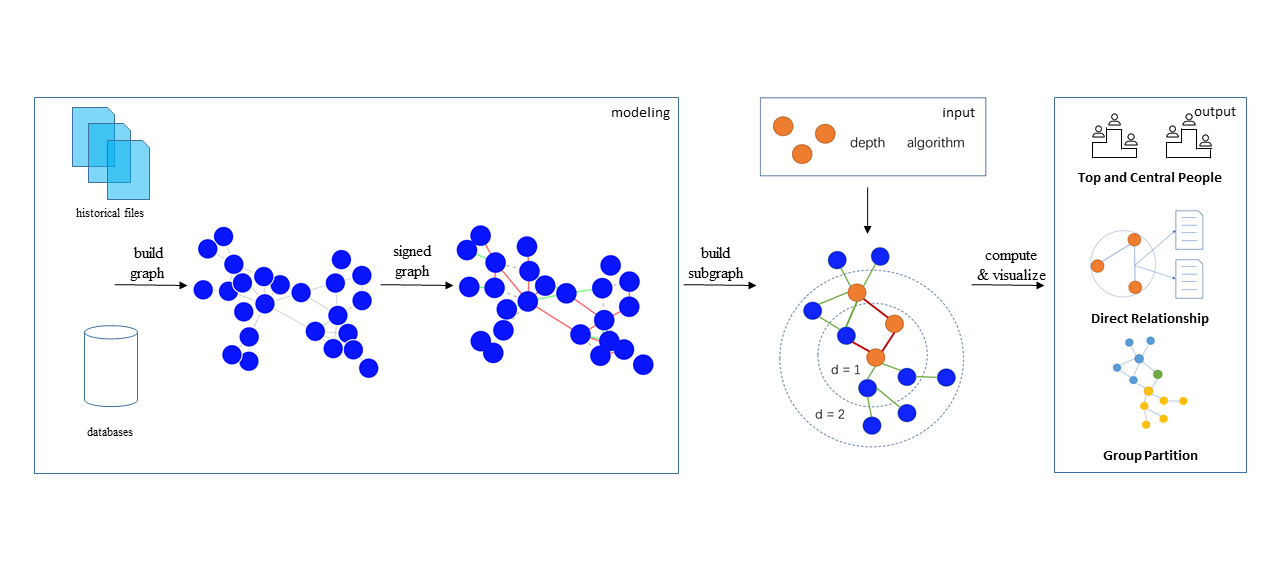}
    \caption{CLHPSoNet Workflow is showed in the Figure. It includes signed graph modeling, subgraph extraction, computing and visualize. The final output is consists of three parts: Top and Central People, Direct Relationship and Group Partition. }
    \label{fig:framework}
\end{figure*}

In the section, we proposed our framework workflow as Figure \ref{fig:framework}.
It includes signed graph modeling, subgraph extraction, computing and visualize. The final output is consists of three parts: Top and Central People, Direct Relationship and Group Partition.

After modeling, we have a signed subgraph and top central people. 
Then we want to know how to parted into different groups based on the social relationship? 
It can be formal defined $CC problem$ as follows\cite{levorato2015ils}:
\begin{theorem}
    (CC problem). Let $G = (V, E, s)$ be a signed graph and we be a nonnegative edge weight associated with each edge $e\in E$. The correlation clustering problem is the problem of finding a partition $P$ of $V$ such that the
    imbalance $I(P)$ is minimized. $I(P)$ is defined as for a partition $P = \{S_1 , S_2 , . . . , S_l \}$:
    $$ I(P) = \sum_{1 \leq i \leq l} \omega^{-}(S_i, S_i) + \sum_{1 \leq i \leq j \leq l} \omega^{+}(S_i, S_j)$$
    where $\omega^{+}(S_i, S_j) = \sum_{e\in E^{+} \cap E[S_i:S_j]} w_e$ and $\omega^{-}(S_i, S_j) = \sum_{e\in E^{-} \cap E[S_i:S_j]} w_e$ 

\end{theorem}

As we said in the related work, the CC problem is NP-hard\cite{doreian1996partitioning} and there are some algorithms can give the solutions.

\subsection{Algorithms}

\subsubsection{Graph Partition}
Doreiana\cite{doreian2009partitioning} proposed a heuristic approach proposed is a simple greedy neighborhood search procedure that assumes a prior knowledge of the number of clusters in the solution. This heuristic is implemented in software Pajek\footnote{http://vlado.fmf.uni-lj.si/pub/networks/pajek/}. It can be a solution algorithm for our problems.

\subsubsection{Community Detection}

Community Detection is developed by the concept known as $modularity$\cite{girvan2002community}. While origin modularity approaches take for granted that links are positively valued, Traag et al\cite{traag2009community} extend an existing model to negative links. 
We use it as one solution of our group partition problems.


\subsubsection{Network Embedding}
Network Embedding is a new method to model the signed graph.
SiNE\cite{wang2017signed} is a deep learning framework for Signed Network Embedding. 
It defines a new objective function guided by the extended structural balance theory. 
The framework has two hidden layers and it will embed the node to $d$-dimension(In his paper, $d=20$ achieved the best performs). 
After the node embedding, it can be used K-means clustering methods to part the nodes to different groups.

Overall, the problem is an open issue to a certain extent. 
Different datasets, research questions, and algorithms will lead to different results. 
There is no particularly good silver bullet.

\section{Application}

\subsection{Development tools}
Based on our framework, we use Python3.5 and JavaScript to build a WebApp for Song Dynasty.
The web backend framework is Flask1.0.1 and the web UI framework is Reactjs16.3.2, antd3.4.4 and D3jsv4. Both our source code and demo system are publicly available online\footnote{https://github.com/huangjunjie95/CLHPSoNet}.

\subsection{A Case Study for Eight Great Prose Masters of Song}

In order to study the relationship between in Eight Great Prose Masters of the Song(Su Shi, Wang Anshi, Ouyang Xiu, Zeng Gong, Su Zhe, Su Xun), we input these people and set $d = 0$, which means just take their relationship into consideration. 
And we choose Community Detection as our algorithm. 
We will get a three-part report by our visualize system.

\subsubsection{Top Central People}
First of all, we can get the central people. 
\begin{figure}[!ht]
    \centering
    \includegraphics[width=0.5\textwidth]{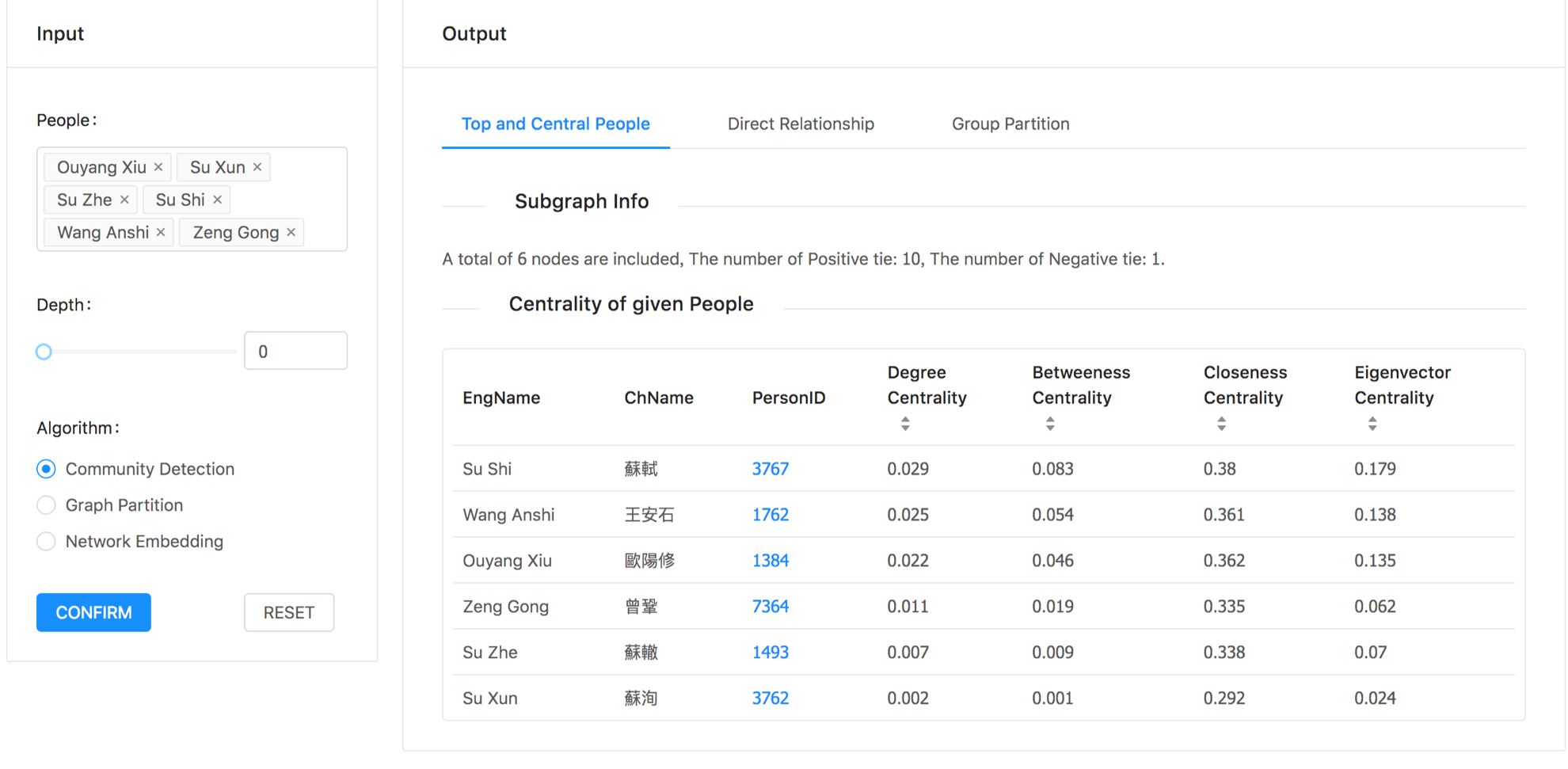}
    \caption{Input: People(Su Shi , Wang Anshi, Ouyang Xiu, Zeng Gong, Su Zhe , Su Xun), Depth(0), Algorithm(Community Detection), Output: People's Centrality Metrics}
    \label{fig:example1}
\end{figure}
Figure \ref{fig:example1} shows that Su Shi, Wang Anshi, Ouyang Xiu all have high central values, indicating that they are important nodes in the network. 
In the social network, many people know each other through them. 
This result meets with common sense. 
These three people's popularity in the understanding of ordinary people is relatively high than other three people.

\subsubsection{Direct Relationship}

The second part reports the direct relationship between given people. 
\begin{figure}[!ht]
    \centering
    \includegraphics[width=0.5\textwidth]{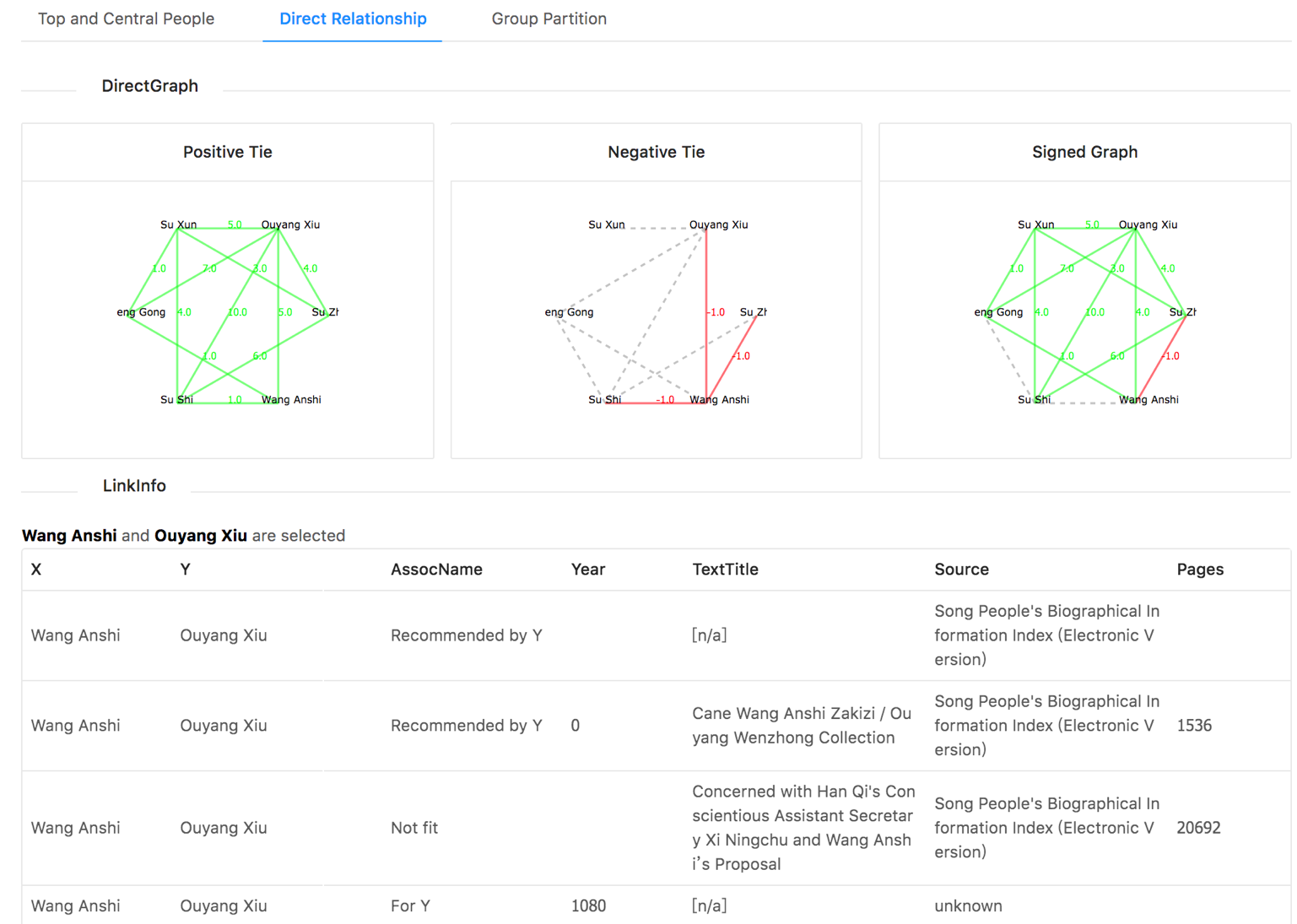}
    \caption{Focus on the Relationship between Wang Anshi and Ouyang Xiu}
    \label{fig:example2}
\end{figure}
We describe in detail the positive relationship, negative relationship and final signed relationship between them. Figure \ref{fig:example2} shows that most of the relationships between them are positive, and the source of the negative relationship is mainly due to political opinion differences. 
It is worth pointing out that the relationship between Ouyang Xiu and Wang Anshi is more complicated.

In the early days, Ouyang Xiu appreciated Wang Anshi’s talent very much. 
He had recommended him to the court several times, and Wang Anshi was also very grateful for the grace of understanding. 
Later, Wang Anshi presided over the reform process. 
Ouyang Xiu's political views are different from Wang. 
The relevant studies have different opinions on the relationship between the two in the later period\cite{guyongxin2001}. 
However, according to our analysis, we believe that the relationship between the two is still positive. After the death of Ouyang Xiu, Wang Anshi also made a memorial service for him. 
With regard to the results of this part, we have found relevant research to support our findings.

\subsubsection{Group Partition}

\begin{figure}[!ht]
    \centering
    \includegraphics[width=0.5\textwidth]{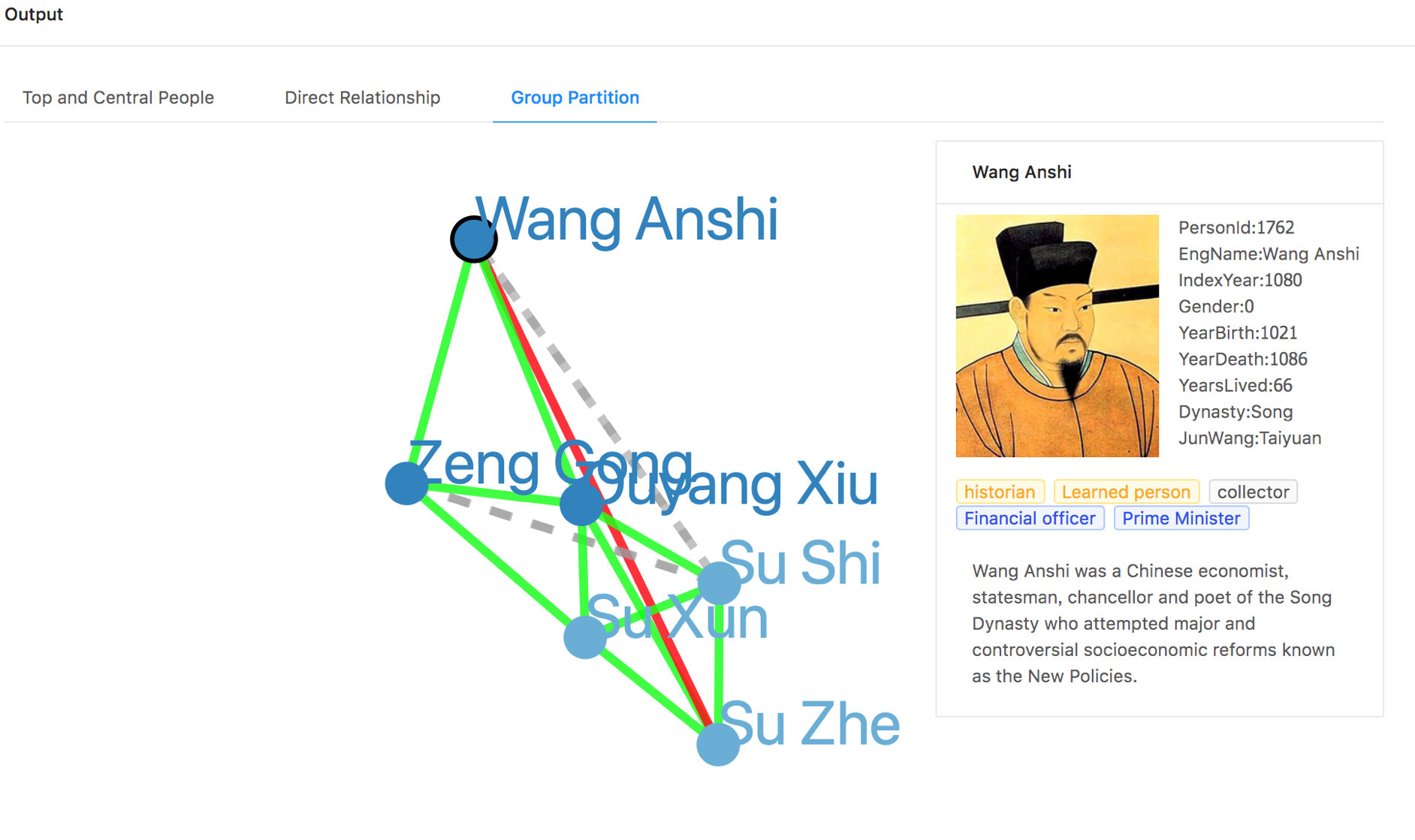}
    \caption{The Partition between Eight Great Prose Masters of Song}
    \label{fig:example4}
\end{figure}

Finally, we obtained the results of the graph partition results in Figure \ref{fig:example4}. 
The system divide Su Shi, Su Zhe, Su Xun into a group, and divide  Wang Anshi, Ouyang Xiu, Zeng Gong into another group. 

In fact, Su Shi, Su Zhe, Su Xun are a family like Alexandre Dumas family and the Brontë family in history. 
Su Xun is the father of Su Shi and Su Zhe. 
Although they are a family, their personality, works, and experience are totally different. 
But if we focus on the six figures networks, it will be found that Su family all opposed Wang's opinions, which matches the historical facts. 
Compare to Su family, Ouyang Xiu, Zeng Gong is more friendly to Wang.
And when you make depth $d \geq 1$, it will give different results. 
The partition of this six people in a large network is not stable, which are not the same like $d=1$.

\section{Conclusion and Future work}
In this paper, we present a new research framework for exploring the social relationship of historical people. 
Based on the proposed the signed social networks model and group partition algorithm, we have built an application to help people analyze and understand the social relationships of the ancients. Via our framework, the social research questions can be transformed into a computing task. People not only can easily use our application to visualize ancient figures' social network but also more clearly understand kinds of literature opinion. 
In the case study, our system produced some information that meets the facts and social experts' judgment. 
There are still several works to be done in future. 
For example, CBDB dataset need to be enriched and improved. 
How to apply machine learning methods to label social network edges is a research problem.
Further study is required for more reasonable and stable for graph partition and structural balance.

\section*{Acknowledgment}
This paper is inspired by Peking University Digital Humanities Forum, and Prof Luo’s Network Science Course.
\bibliographystyle{IEEEtran}
\bibliography{IEEEabrv,references}
\end{document}